\documentclass{elsart}

 \usepackage{graphicx}

\begin{document}

\begin{frontmatter}



\title{Elementary vortex pinning potential in superconductors with
unconventional order parameter}


\author[okayama]{Nobuhiko Hayashi\corauthref{cor1}}
\ead{hayashi@mp.okayama-u.ac.jp}
and
\author[tokyo]{Yusuke Kato}

\address[okayama]{Computer Center, Okayama University,
Okayama 700-8530, Japan}
\address[tokyo]{Department of Basic Science, University of Tokyo,
Tokyo 153-8902, Japan}

\corauth[cor1]{Corresponding author. Fax: +81-86-251-7244}

\begin{abstract}
   The elementary vortex pinning potential is studied
in unconventional superconductors
within the framework of the quasiclassical theory of superconductivity.
   Numerical results are presented
for $d$-, anisotropic $s$-, and isotropic $s$-wave superconductors
to show explicitly that
in unconventional superconductors
the vortex pinning potential is determined
mainly by the loss of the condensation energy in bulk due to
the presence of
the pinning center, i.e., by the breakdown of the Anderson's theorem.
   It is found that the vortex pinning energy
in the $d$-wave pairing case
is 4 -- 13 times larger than those in the $s$-wave pairing cases.
This means that an enhancement of pinning effect in unconventional
superconductors occurs due to the breakdown of the Anderson's theorem.
   The case of a chiral $p$-wave superconductor is also
investigated in terms of the vortex core states subject to
the Andreev reflection, where important is whether
the vorticity and chirality are parallel or antiparallel.
\end{abstract}

\end{frontmatter}

\section{Introduction}
\label{intro}
   Much attention has been focused on the vortex pinning
in type-II superconductors.
   The vortex pinning in the superconductors under magnetic fields
plays an important role on various vortex-related quantities and phenomena
such as the critical current,
hysteresis of the magnetization, and quantum vortex tunneling.
   The problem of the vortex pinning is categorized
in two aspects.
   One is the elementary vortex pinning force, which is
the interaction between a vortex and a single defect.
   The other is the summation problem, which is
how the elementary vortex pinning forces add up to a holding
force on the elastic vortex lattice under a distribution
of defects \cite{Kerchner83}.
   A study of the elementary vortex pinning is the step
necessary for the understanding of the phenomena
related to the vortex pinning in the superconductors
under magnetic fields.

   Traditionally, the mechanism of the elementary vortex pinning
is considered to be that a defect locally prohibits superconducting
condensation so that it attracts a normal region of a vortex core
in order to avoid the loss of the condensation energy.
   Thuneberg {\it et al.} \cite{Thuneberg82,Thuneberg84}
advanced the understanding of the mechanism of
the elementary vortex pinning,
taking account of a nonlocal effect
that a defect as a scattering center scatters the quasiparticles.
   This mechanism of the vortex pinning is that
the quasiparticle scattering by the defect
inside a vortex core
helps to avoid suppressions of
the order parameter around the vortex center
up to distances of the order of the coherence length.

   In this paper, we investigate the elementary vortex pinning
potential in unconventional superconductors,
following the analysis started by
Thuneberg {\it et al.} \cite{Thuneberg82,Thuneberg84}
for conventional $s$-wave superconductors.
   The unconventional superconductivity is recently
proposed for many superconductors
such as high-$T_{\rm c}$ cuprates, organic conductors, and
heavy-fermion compounds.
   For example, the $d$-wave superconductivity is believed to
be realized in most of the high-$T_{\rm c}$ cuprates,
and the chiral $p$-wave one is expected in
a non-copper-layered perovskite superconductor
Sr$_2$RuO$_4$ \cite{Sigrist99}
(although it seems to be open to
further discussion \cite{lebed00}).
   We calculate numerically the elementary vortex pinning potential
for $d$-, anisotropic $s$-, and isotropic $s$-wave pairings,
and compare them each other to clarify what could be one of
the characteristic
properties specific to the superconductors with
the unconventional order parameter.
   We also present an analytical consideration of
a novel chirality-dependent vortex pinning
in a chiral $p$-wave superconductor.

\section{Formulation}
\label{intro}
   To investigate the vortex pinning, we use
the quasiclassical theory of superconductivity \cite{serene}.
   We start with the Eilenberger equation for
the quasiclassical Green function
in the absence of the pinning,
%
\begin{equation}
{\hat g}_{\rm imt}(i\omega_n,{\bf r},{\bar{\bf k}})=
-i\pi
\pmatrix{
g_{\rm imt} &
if_{\rm imt} \cr
-if^{\dagger}_{\rm imt} &
-g_{\rm imt} \cr
},
\label{eq:qcg}
\end{equation}
namely,
%
\begin{eqnarray}
i v_{\rm F} {\bar{\bf k}} \cdot
{\bf \nabla}{\hat g}_{\rm imt}
+ \bigl[ i\omega_n {\hat \tau}_{3}-{\hat \Delta},
{\hat g}_{\rm imt} \bigr]
=0,
\label{eq:eilen}
\end{eqnarray}
where ${\hat \tau}_{3}$ is the Pauli matrix and the order parameter
\begin{eqnarray}
{\hat \Delta}({\bf r},{\bar{\bf k}}) =
\pmatrix{0 & \Delta({\bf r},{\bar{\bf k}})\cr
-\Delta^*({\bf r},{\bar{\bf k}})
& 0 \cr}.
\label{eq:matrix-a}
\end{eqnarray}
The Eilenberger equation (\ref{eq:eilen})
is supplemented by the normalization condition
%
$
{\hat g}_{\rm imt}(i\omega_n,{\bf r},{\bar{\bf k}})^2
=-\pi^2{\hat 1}
$,
and the commutator is
$[{\hat a},{\hat b}]={\hat a}{\hat b}-{\hat b}{\hat a}$.
   The symbol ${\bf r}=(r\cos\phi,r\sin\phi)$ denotes
the center of mass coordinate of the Cooper pairs
and
${\bar{\bf k}}=(\cos\theta,\sin\theta)$ denotes
the relative coordinate
of them.
   The cylindrical Fermi surface is assumed.
   We use units in which $\hbar = k_{\rm B} = 1$.

   Following
Thuneberg {\it et al.} \cite{Thuneberg82,Thuneberg84,Thuneberg81},
the effect of the pinning is introduced as follows to
the quasiclassical theory of superconductivity.
   The quasiclassical Green function ${\hat g}$
in the presence of a point-like non-magnetic defect situated at
${\bf r}={\bf R}$
is obtained from the Eilenberger equation
\begin{eqnarray}
i v_{\rm F} {\bar{\bf k}} \cdot
{\bf \nabla}{\hat g}
+ \bigl[ i\omega_n {\hat \tau}_{3}-{\hat \Delta}, {\hat g} \bigr]
=\bigl[ {\hat t}, {\hat g}_{\rm imt} \bigr] \delta ({\bf r}'),
\label{eq:eilen-pin}
\end{eqnarray}
and the $t$ matrix due to the defect
\begin{eqnarray}
{\hat t}(i\omega_n, {\bf r}') =
\frac{v}{D} \Bigl[
{\hat 1} + N_{0} v
\langle {\hat g}_{\rm imt}(i\omega_n, {\bf r}',{\bar {\bf k}}) \rangle_\theta
\Bigr],
\label{eq:t-matrix}
\end{eqnarray}
where ${\bf r}'={\bf r}-{\bf R}$,
the denominator $D=1+(\pi N_0 v)^2 \bigl[
\langle g_{\rm imt} \rangle_\theta ^2
+ \langle f_{\rm imt} \rangle_\theta  \langle f^{\dagger}_{\rm imt} \rangle_\theta
\bigr]$,
the symbol
$\langle \cdots \rangle_\theta = \int \cdots d \theta/2\pi$,
the normal-state density of states on the Fermi surface is $N_{0}$,
and
we assume the $s$-wave scattering $v$ to obtain Eq.\ (\ref{eq:t-matrix}).
The cross section of the defect is given by
$\sigma_{\rm tr} =(4\pi/k_{\rm F}^2)\sin^2 \delta_0$
with the scattering phase shift $\delta_0$,
and
$\tan \delta_0 = -\pi N_0 v$.

   The free energy in the presence of the defect is,
at the temperature $T$, given as
\cite{Thuneberg82,Thuneberg84,Thuneberg81}
\begin{eqnarray}
\delta \Omega ({\bf R}) =
N_0 T \int^{1}_{0} d \lambda
\sum_{\omega_{n}}
\int d {\bar {\bf k}}
\int d {\bf r}
{\rm Tr}
\bigl[
\delta {\hat g}_{\lambda} {\hat \Delta}_b
\bigr],
\label{eq:free-ene}
\end{eqnarray}
where $\delta {\hat g}_{\lambda} = {\hat g}-{\hat g}_{\rm imt}$
is evaluated at
${\hat \Delta}= \lambda {\hat \Delta}_{b}$,
and
${\hat \Delta}_{b}$
is the order parameter in the absence of the defect.
Equation (\ref{eq:free-ene}) represents the difference
in the free energy
between the states with and without the defect, and then
gives the vortex pinning potential $\delta \Omega ({\bf R})$.

\begin{figure}
\includegraphics[width=8cm]{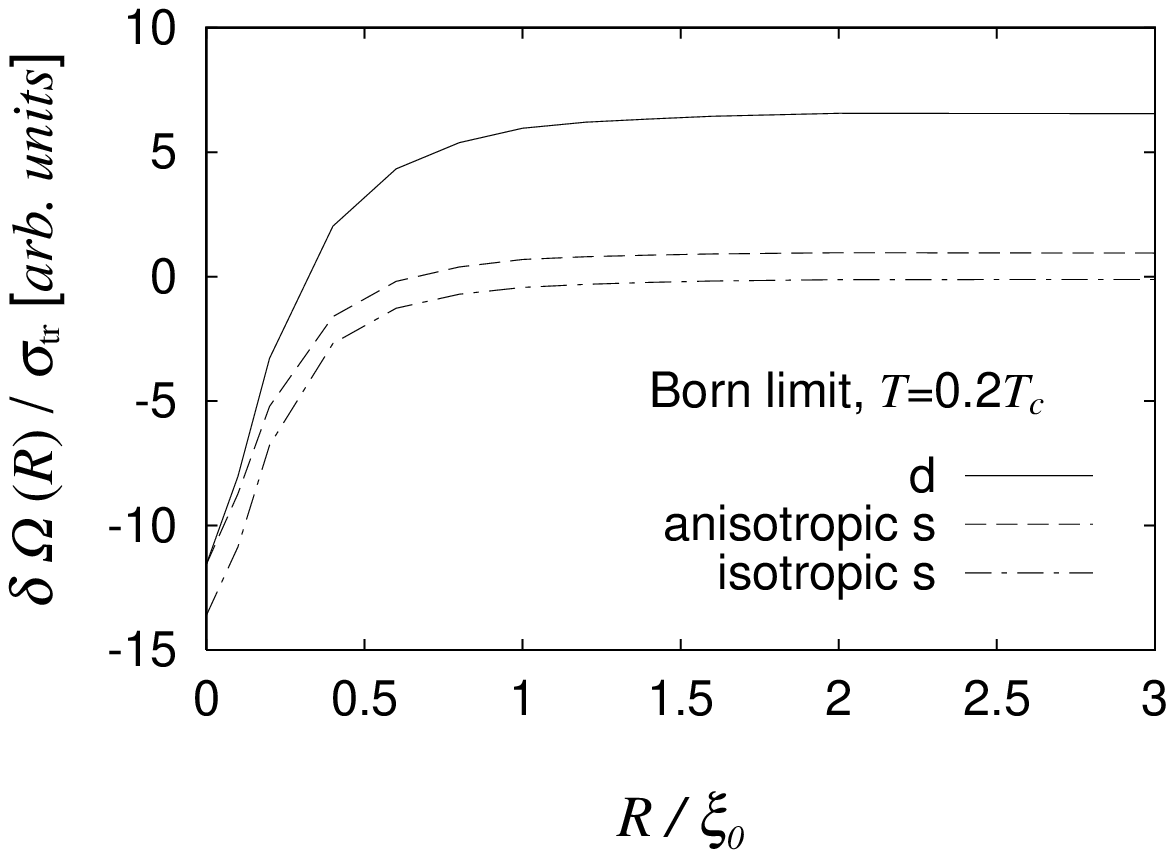}
\includegraphics[width=8cm]{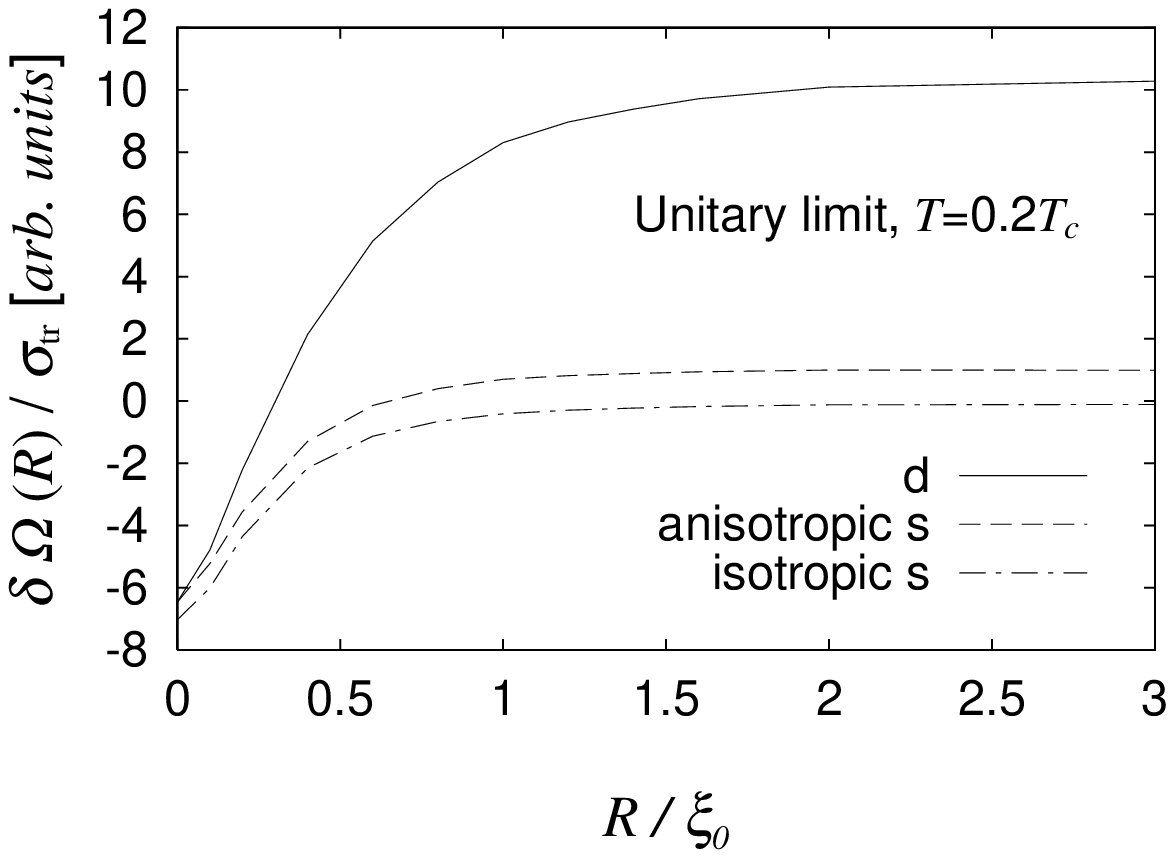}
\includegraphics[width=8cm]{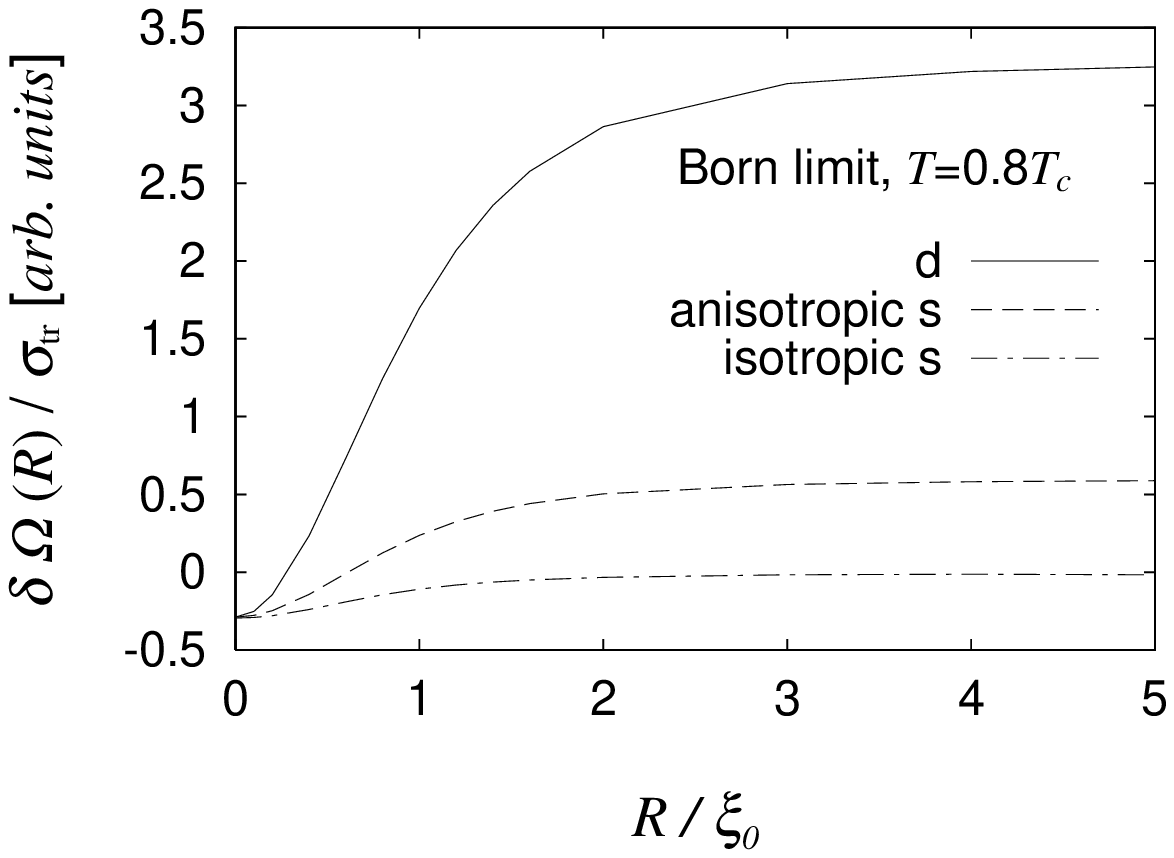}
\includegraphics[width=8cm]{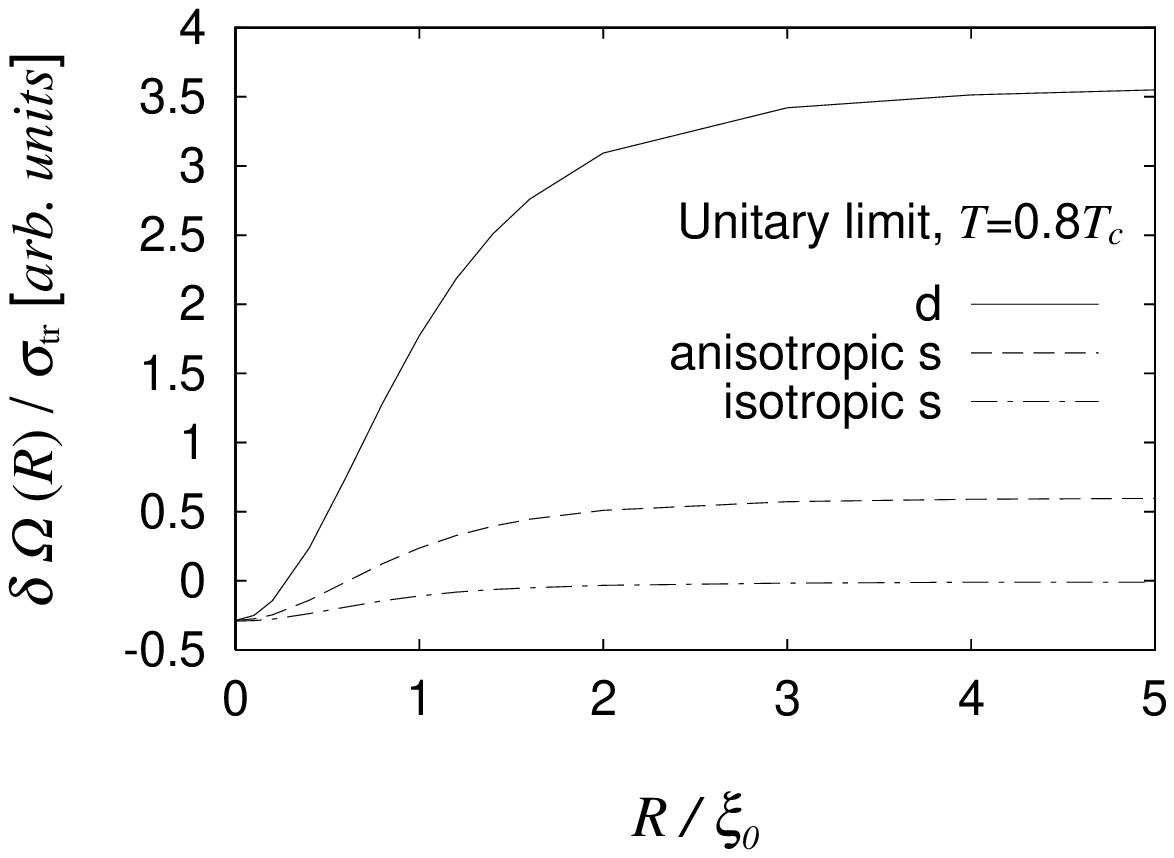}
\caption{
   The vortex pinning potential as a function of
the distance $R$ between the vortex center and the defect.
The defect is situated on the $x$ axis (in the direction of $\phi=0$).
}
\label{fig:1}
\end{figure}

\section{Results}
\label{results}
\subsection{$d$-, anisotropic $s$-, and isotropic $s$-wave pairings}
\label{subsec:d-wave}
   We calculate the vortex pinning potential $\delta \Omega ({\bf R})$
numerically with the order parameter around a vortex,
$\Delta_{b}({\bf r},{\bar {\bf k}})=F(\theta) \Delta_{\rm BCS}(T)
\tanh \bigl[ r/\xi(T) \bigr] \exp(i\phi)$,
(note that the vortex center is situated at ${\bf r}=0$).
   Here,
the ${\bar {\bf k}}$-dependent function
$F(\theta)=1$ for isotropic $s$-wave,
$\sqrt2 |\cos2\theta|$ for anisotropic $s$-wave,
and $\sqrt2 \cos2\theta$ for $d$-wave pairings.
   $\Delta_{\rm BCS}(T)$ is the BCS order parameter amplitude and
$\xi(T)$ the coherence length.
$\bigl(
\xi(T)=\xi_0 \cdot \bigl[ \Delta_0 / \Delta_{\rm BCS}(T) \bigr]$,
$\Delta_0=\Delta_{\rm BCS}(0),$ and $\xi_0= v_{\rm F} / \Delta_0.
\bigr)$

   In Fig.\ \ref{fig:1},
we show the numerical results for the Born limit ($\delta_0 \rightarrow 0$)
and the unitary limit ($\delta_0 = \pi/2$).
   The defect is situated on the $x$ axis (in the direction of $\phi=0$).
   In the case of the isotropic $s$-wave pairing,
when the defect is situated in bulk, i.e., far away from the vortex center
($R=|{\bf R}| \rightarrow \infty$),
the difference in the free energy
between the states with and without the defect,
$\delta \Omega (R)$,
is equal to zero.
   This is because the Anderson's theorem \cite{Anderson59}
is valid
for homogeneous systems
in isotropic $s$-wave superconductors
and the non-magnetic defect has no influence on the free energy.
   On the other hand, in the anisotropic pairing cases,
$\delta \Omega (R \rightarrow \infty)$ is finite and positive
as seen in Fig.\ \ref{fig:1},
which is the loss of the condensation energy in bulk
due to the presence of the defect and
to the breakdown of the Anderson's theorem.
   The condensation energy loss in bulk
$\delta \Omega (R \rightarrow \infty)$
contributes to
the depth of the vortex pinning potential $\delta \Omega (R)$,
i.e., the vortex pinning energy.
   While the contribution of $\delta \Omega (R \rightarrow \infty)$
to the vortex pinning energy was first
pointed out by Thuneberg {\it et al.} \cite{Thuneberg81}
in the context of an ion in $p$-wave superfluid $^3$He-B,
we performed here
a numerical calculation explicitly for the vortex pinning potential in
the $d$-wave superconductors.
   It is noted from Figs.\ \ref{fig:1}(c) and \ref{fig:1}(d) that,
at $T=0.8T_{\rm c}$ (high temperature),
the vortex pinning energy in the $d$-wave case
is dominantly determined by
the condensation energy loss in bulk
$\delta \Omega (R \rightarrow \infty)$,
and is
4 -- 13 times larger than those in the $s$-wave cases.

%
\subsection{chiral $p$-wave pairing}
\label{subsec:p-wave}
   For the chiral $p$-wave pairing
${\bf d}={\bf z}({\bar k}_x \pm i{\bar k}_y)$
\cite{Sigrist99},
we analytically investigate the vortex pinning potential.
   In the case of that chiral $p$-wave pairing,
there exist
the following two possible forms for
the order parameter
$\Delta(r,\phi;\theta)$ $\bigl( =\Delta({\bf r},{\bar {\bf k}})\bigr)$
of axisymmetric vortex
\cite{Heeb99,Matsumoto01,Kato01}:
(i)
$\Delta_b^{+-}(r,\phi;\theta)=
\Delta_{+}(r) \exp\bigl[i(\theta-\phi)\bigr]
+ \Delta_{-}(r) \exp\bigl[i(-\theta+\phi)\bigr]$,
where the chirality and vorticity are antiparallel;
(ii)
$\Delta_b^{++}(r,\phi;\theta)=
\Delta_{+}(r) \exp\bigl[i(\theta+\phi)\bigr]
+ \Delta_{-}(r) \exp\bigl[i(-\theta+3\phi)\bigr]$,
where the chirality and vorticity are parallel.
   Here, $\Delta_{\pm}(r)$ are real,
$\Delta_{+}(r \rightarrow \infty)=\Delta_{\rm BCS}(T)$,
and $\Delta_{-}(r \rightarrow \infty)=0$.

   We focus on $\delta \Omega(R=0)$, where
both the defect and the vortex center are situated just at the origin
${\bf r}=0$.
   The quasiparticles inside the vortex core,
subject to the Andreev reflection,
run along straight lines, namely
the quasiclassical trajectories \cite{Kato01,Kato00}.
   We consider the quasiclassical trajectories
which go through the origin ${\bf r}=0$.
   On those trajectories with zero impact parameter,
the position vector is parallel to the direction of the trajectory
(i.e., ${\bf r} \parallel {\bar {\bf k}}$), and therefore
$\phi=\theta,\ \theta+\pi$.
   In this situation the order parameter is
$\Delta_b^{+-}(r, \phi; \theta) =\pm
\bigl( \Delta_{+}(r) + \Delta_{-}(r) \bigr)$
in the case of (i),
and
$\Delta_b^{++}(r, \phi; \theta) =\pm
\bigl( \Delta_{+}(r) + \Delta_{-}(r) \bigr)
\exp(2i\theta)$
in the case of (ii).
   Of importance is the difference in the phase factor
of these order parameters of (i) and (ii).

   On the basis of an analysis of
the so-called zero-core vortex model in Ref.\ \cite{Thuneberg84},
the matrix elements of
${\hat g}_{\rm imt}$ at the vortex center are approximately obtained as
$g_{\rm imt}=\sqrt{\omega_n^2 +|{\tilde \Delta}|^2} /\omega_n$,
$f_{\rm imt}=-{\tilde \Delta} / \omega_n$, and
$f^{\dagger}_{\rm imt}= {\tilde \Delta}^{*} / \omega_n$, where
${\tilde \Delta}=
\Delta_b^{+\pm}(r \rightarrow \infty, \phi=\theta; \theta)$.
   Because of the difference in the phase factor of the order parameters
mentioned above,
$\langle f_{\rm imt} \rangle_\theta = f_{\rm imt}$ and
$\langle f_{\rm imt}^{\dagger} \rangle_\theta = f_{\rm imt}^{\dagger}$
in the case of (i),
while
$\langle f_{\rm imt} \rangle_\theta = 0$ and
$\langle f_{\rm imt}^{\dagger} \rangle_\theta = 0$
in the case of (ii).
   Therefore, in the case of (i),
$\langle {\hat g}_{\rm imt} \rangle_\theta  =  {\hat g}_{\rm imt}$,
and then
$[{\hat t},{\hat g}_{\rm imt}]=0$ from Eq.\ (\ref{eq:t-matrix}).
   In the case of (ii),
$\langle {\hat g}_{\rm imt} \rangle_\theta  \neq  {\hat g}_{\rm imt}$,
and then
$[{\hat t},{\hat g}_{\rm imt}] \neq 0$.

   When $[{\hat t},{\hat g}_{\rm imt}]=0$,
the Eilenberger equation (\ref{eq:eilen-pin})
in the presence of the defect
is identical to Eq.\ (\ref{eq:eilen})
(the equation in the absence of the defect),
i.e., the defect has no influence on the Green function and
the free energy.
   From it and the above analysis,
$\delta \Omega(0) = 0$ in the case of (i)
when the chirality is antiparallel to the vorticity,
and $\delta \Omega(0) \neq 0$ in the case of (ii)
when the sense of the chirality is the same as that of the vorticity.
   It means that the vortex pinning depends on the chirality
in the chiral $p$-wave superconductor.
   This analytical result is certainly confirmed by a numerical
calculation \cite{Hayashi} with self-consistently obtained
order parameters
$\Delta_b^{+\pm}(r,\phi;\theta)$ \cite{Kato01}.

\section{Summary}
\label{summary}
   We investigated the elementary vortex pinning potential
$\delta \Omega (R)$
on the basis of the quasiclassical theory of superconductivity.
   The numerical results were presented in Fig.\ \ref{fig:1}
for the $d$-wave and $s$-wave pairings.
   The vortex pinning energy in the $d$-wave pairing case
is about 10 times larger than those in the $s$-wave pairing cases,
at a high temperature, because of the breakdown of the Anderson's theorem
in unconventional superconductors.
   We also investigated analytically the quasiparticles inside the vortex core
in the chiral $p$-wave superconductor,
and found the chirality-dependent vortex pinning.
   We expect that the chirality-dependent vortex pinning
revealed in this paper
shares the same physics with
impurity effects on
the quasiparticle relaxation time \cite{Kato00}
and on the energy spectrum \cite{Volovik,Matsumoto}
inside the vortex core in chiral $p$-wave superconductors.

\begin{ack}
One of the authors (N.H.)
thanks M.\ Ichioka, K.\ Machida, M.\ Takigawa,
N. Nakai, and M.\ Matsumoto for useful discussions.
   Y.K. thanks M.\ Sigrist for useful discussions.
   This work is partly supported by
Grant-in-Aid for Scientific Research on Priority Areas (A)
of ^^ ^^ Novel Quantum Phenomena in Transition Metal Oxides" (12046225)
from the Ministry of Education, Science, Sports and Culture and
Grant-in-Aid for Encouragement of Young Scientists from Japan Society for
the Promotion of Science (12740203).
\end{ack}





\end{document}